\documentclass[%
 aip,
 amsmath,amssymb,
 reprint,%
]{revtex4-1}
\usepackage{graphicx}
\usepackage{amssymb}
\usepackage{amsmath}
\usepackage{amsfonts}
\usepackage{mathrsfs}
\usepackage{color}
\usepackage{physics}
\usepackage[english]{babel}
\usepackage[utf8]{inputenc}
\usepackage[T1]{fontenc}
\usepackage{mathptmx}
\usepackage{etoolbox}

\makeatletter
\def\@email#1#2{%
 \endgroup
 \patchcmd{\titleblock@produce}
  {\frontmatter@RRAPformat}
  {\frontmatter@RRAPformat{\produce@RRAP{*#1\href{mailto:#2}{#2}}}\frontmatter@RRAPformat}
  {}{}
}%
\makeatother

\begin{document}

\preprint{AIP/123-QED}

\newcommand{\half}{\frac{1}{2}}
\newcommand{\pcl}[1]{#1_{\mathrm{p}}}

\title[A classical analog of  the quantum Zeeman effect ]{A classical analog of  the quantum Zeeman effect}
\author{P. Jamet}
\email{pierre.jamet@neel.cnrs.fr} 
 \affiliation{Univ. Grenoble Alpes, CNRS, Grenoble INP, Institut Neel, F-38000 Grenoble, France}
\author{A. Drezet}%
 \email{aurelien.drezet@neel.cnrs.fr}
\affiliation{Univ. Grenoble Alpes, CNRS, Grenoble INP, Institut Neel, F-38000 Grenoble, France}

\begin{abstract}
    We extend a recent  classical mechanical analog of Bohr's atom consisting of a scalar field coupled to a massive point-like particle [P. Jamet, A. Drezet, ``A mechanical analog of Bohr's atom based on de Broglie's double-solution approach'', Chaos \textbf{31}, 103120 (2021)] by adding and studying the contribution of a uniform weak magnetic field on their dynamics. In doing so, we are able to recover the splitting of the energy levels of the atom called Zeeman's effect within the constraints of our model and in agreement with the semiclassical theory of Sommerfeld. This result is obtained using Larmor's theorem for both the field and the particle, associating magnetic effects with inertial Coriolis forces in a rotating frame of reference. Our work,  based on the old `double solution' theory of de Broglie, shows that a dualistic model involving a particle guided by a scalar field can reproduce the normal Zeeman effect. 
\end{abstract}

\date{\today}

\maketitle
\section*{Lead Paragraph}
\textbf{Quantum mechanics is generally opposed to classical physics. However, Louis de Broglie already attempted, and partially succeeded, in giving a physical spatio-temporal interpretation of wave-particle duality in the 1920's. In the present work we further develop a classical analog model of the old Bohr-Sommerfeld atom involving a particle coupled to a wave. Here, we introduce a coupling with an external magnetic field and show, using the Larmor theorem, that we can recover the famous Zeeman splitting of energy levels predicted in the old quantum theory.  This we believe can shed new light on the relations between the quantum and the classical worlds an will provide new ways for observing this relation experimentally.}
\section{Introduction}
As early as 1897\cite{Zeeman}, and following the interests of Michael Faraday that led him to perform his last experiments on this subject in 1862\cite{BenceJones1870}, Pieter Zeeman published a study on the effects of a magnetic field on the characteristics of the light emitted by various sources. 
He observed at the time that the spectral lines of sodium salts were widened when the substance was placed near a magnetic field.
With his experiments he was able to not only identify this widening as being related only to the microscopic properties of the sources and not on the thermodynamic state of the substance, but also to explain its origin using Hendrik Lorentz's theory\cite{Lorentz} with whom he shared the Nobel prize in 1902. 
The theoretical explanation at the time was that light was emitted by oscillating or orbiting discrete charges, so that their orbital period was increased or decreased through the effect of the magnetic field, and in turn the frequency of the light would also be modified accordingly.

\indent Years later in 1913, Niels Bohr proposed a new planetary model\cite{Bohr1913} for the atom, following more than a decade of similar works on the structure of matter -- in particular by Joseph Larmor who wrote himself a planetary model with elliptical orbits explaning Zeeman's observations in late 1897\cite{Larmor1897} -- introducing the hypothesis of quantization for orbital momenta. 
In this model, negatively charged electrons orbit around a positively charged nucleus -- at the time only moving along circular paths but extended in 1916 by Arnold Sommerfeld to elliptical ones\cite{Sommerfeld} -- but only at fixed radii and energies. 
In this theory, light is emitted when an electron jumps from one energy level to a lower one, emitting a quantum of light with frequency equal to the difference in the energies of the two orbits divided by Planck's constant $h$. 
This is clearly reminiscent of Lorentz, Zeeman, and Larmor's explanations of 1897\cite{Lorentz,Zeeman,Larmor1897}, and indeed one can easily recover the same phenomenon of the effect of a magnetic field on the frequency of the elecromagnetic waves, relating the orbital period or the angular frequency to that of the waves. 
Of course this model also has the advantage of explaining the origin of the line structure in the spectra of atoms, and predicts that it is not a widening of these lines but rather a splitting of a single line into $2n + 1$ where $n$ is the orbital angular momentum quantum number characterizing the orbit $L = \oint p \dd{q} = n h$. This explanation, however, could not account for all observations. In fact two phenomena existed, the normal Zeeman effect where a spectral line splits into an odd number of subsequent lines, correctly explained by the theory, and an anomalous one with a line dividing into an even number, as yet unexplained.
Later still, the development of modern quantum mechanics by de Broglie, Schr\"odinger, Heisenberg, Born and many others led to yet another atomic model which once again predicted this normal Zeeman splitting, but also the anomalous one (related to the existence of the half-integer spin for the electron) characterizing any electronic orbital by a set of four quantum numbers $\ket{n, l, l_z, s_z}$\cite{Sommerfeld}.\\ 
\indent More recently, experiments have been performed\cite{Couder2006,Couder2009,Fort2010,Harris2013} on droplets bouncing on a vibrating liquid bath showing that such droplets can exhibit quantum-like properties (for reviews see Refs. \cite{Bush2015a,Bush2015b, Bush2018, Bush2020}). In particular, if the bath itself is rotated, the quantization of the orbital angular momentum of the droplet, induces in turn a quantization of other quantities such as the radii of the orbits, which were shown to follow the same rules as Zeeman or Landau levels\cite{Fort2010}. This phenomenon can be understood as an interaction between the orbital angular momentum of the droplet and an ``effective magnetic field'' which actually is an inertial force arising from the rotating reference frame.\\
\indent In this paper, we will go back to the old quantum theory and start from a new atomic model developped recently \cite{Drezet2020,Jamet2021} with a mechanical approach related to ideas introduced by Louis de Broglie\cite{debroglie1925,debroglie1927,deBroglie1960},  and add to it an external magnetic field to recover the normal Zeeman splitting of the energy levels. As we will see, this model will indeed explain this phenomenon both from a particle and a field point of view, while also gaining insight into some limitations or constraints on such a procedure. After briefly recalling  in Sec. \ref{sec:1} the characteristics of our previously developped model\cite{Drezet2020,Jamet2021}, we will in particular look first at the dynamics of the particle in Sec. \ref{sec:2a} and then of the guiding field in Sec. \ref{sec:2b}, making use of an interesting method for finding the solutions of complex field equations coupled to a magnetic field using Larmor's theorem\cite{Larmor1897,Larmor}. Moreover, we stress that the general methodology proposed in this work  is  based on  mechanical analogies that can be applied to several fields of physics including acoustic and optics. We discuss the fundamental and applicative potentialities of our work in the final discussion.



\section{Description of our model}
\label{sec:1}
In a previous paper~\cite{Jamet2021} (here after called JD), we developped a classical analog of a quantum  atom based on some ideas first introduced by Louis de Broglie, known as the `double solution', which successfully reproduced the usual quantization rules of Bohr and Sommerfeld. 
More specifically, this model uses a complex scalar field $u(x)$ defining a physical object in space-time (to be distinguished from the usual wave-function $\Psi(x)$ introduced in quantum mechanics) that is coupled to a particle represented by a point-like mass $m_\mathrm{p}$ \textit{via} a holonomic constraint between the field itself and an internal oscillation $z(\tau)$ for the particle ($\tau$ is a relativistic proper time for the particle). 
We also introduced a four-vector potential $A(x)$ to take into account external electromagnetic forces on both the field and the particle, such as a central electrostatic energy potential $U(r)=eV(r) = -\alpha/r$ ($\alpha=e^2/(4\pi)$ denotes Sommerfeld's fine structure constant and we use the convention $e=-|e|<0$). 
Here, we will study this system further by adding an external uniform magnetic field $\vb{B} = B \vu{e}_z$ in order to derive the quantized splitting of the energy levels with respect to the projection of the angular momentum $L_z$, called the Zeeman effect.

As shown in JD, our system is characterized by the action 

\begin{equation}
    \begin{split}
		I = &-\int\bqty{\pcl{m} - \half m_{\mathrm{p}}\sigma\pqty{\vqty{\dot{z}(\tau)}^2 - \pcl{\Omega}^2 \vqty{z(\tau)}^2}}\dd\tau \\
		&+ \int \left\lbrace \mathcal{N}(\tau)\bqty{z(\tau) - u(\pcl{x}(\tau))}^*\right. \\&+ \left. \mathcal{N}^*(\tau)\bqty{z(\tau) - u(\pcl{x}(\tau))}\right\rbrace\dd\tau \\ 
		&- e\int A(\pcl{x}(\tau))\pcl{\dot{x}}(\tau)\dd\tau + T\int(Du)(Du)^*\dd^4 x 
	\end{split}
	\label{eq:action}
\end{equation}
which contains several physical quantities listed in Table~\ref{table1} and discussed in more details in JD. In our work we use natural units ($c = \hbar = 1$ and the Minkowski metric $\eta_{\mu\nu}$ with signature $(1,-1,-1,-1)$.\\
\begin{table}[h]
\begin{tabular}{l||c}
 Parameters & Physical meaning \\
\hline
\hline
$u(x)$  & \textrm{Fundamental field}\\
\hline
$A^\mu(x)$  & \textrm{Electromagnetic 4-vector potential}\\
\hline
$z(\tau)$  & \textrm{Internal oscillator} \\
\hline
$\mathcal{N}(\tau))$  & \textrm{Internal reaction force acting upon the particle} \\
\hline
$\pcl{x}^\mu(\tau)$  & \textrm{4-vector position of the particle} \\
\hline
$T$  & \textrm{Tension of the field} \\
\hline
$\pcl{\Omega}$  & \textrm{Internal oscillator pulsation} \\
\hline
$m_{\mathrm{p}}$  & \textrm{Bare particle mass} \\
\hline
$\sigma$  & \textrm{Coupling constant} \\
\hline
$e=-|e|$  & \textrm{Particle electric charge} \\
\end{tabular}
\caption{Table summarizing the different parameters  of the model. }\label{table1}
\end{table}
\indent   The physical meaning of the different contributions to Eq.~\ref{eq:action} must be briefly discussed.  The first line in Eq.~\ref{eq:action} contains a single particle relativistic Lagrangian where $\dd\tau=\sqrt{\eta_{\mu\nu}dx_\mathrm{p}^\mu(\tau)dx_\mathrm{p}^\nu(\tau)}$ is a Minkowski proper time interval defined along the particle trajectory (repeated indices are implicitly summed over), and $m_\mathrm{p}$ is a rest mass for the particle with trajectory $x_\mathrm{p}(\tau)$ . The specificity of our model is to generally have  a non constant mass obtained by adding a harmonic oscillator term in the first line of  Eq.~\ref{eq:action}.  This harmonic oscillator is associated with an internal degree of freedom $z(\tau)$ (representing an internal clock for the particle) and a typical pulsation $\pcl{\Omega}$. This term plays a key role in the analogy with de Broglie's theory as we show below.  We mention that the addition of this term generalizes our non-relativistic mechanical analogy developed in~\cite{Drezet2020}.     The second and third lines of the action Eq.~\ref{eq:action} define a holonomic constraint connecting the wave $u(x)$ and the particle at the position $x=x_\mathrm{p}(\tau)$.  The terms  $\mathcal{N}(\tau)$, $\mathcal{N}^\ast(\tau)$  play the role of reaction forces between the wave, the internal clock, and the particle.  Finally, the last line in   Eq.~\ref{eq:action}  describes a standard quadratic field Lagrangian density for a d'Alembert scalar wave equation ($T$ is a constant similar to a tension in a elastic string~\cite{Drezet2020}). Here, however the field $u(x)$ is complex valued and instead of the usual partial derivative $\partial_\mu$ we have a covariant derivative $D_\mu = \partial_\mu + i e A_\mu$ depending on the external electromagnetic potential $A(x)$ interacting with the $u-$wave.  The model is thus gauge invariant in a way similar to the Klein-Gordon equation. The Last line of Eq.~\ref{eq:action}  also includes a standard coupling term for the particle and the external electromagnetic potential $A_\mu(x)$.   We stress that here we have used the same electric charge $e$ for both the particle and the covariant derivative $D_\mu$.  As explained in JD this hypothesis could be relaxed but will be considered in the following.       \\   
\indent Using a variational principle $\delta I=0$, Eq.~\ref{eq:action} leads to a system of coupled differential equations, from which we can extract a regime called transparency~\cite{Jamet2021,Drezet2020,Borghesi2017} where the force $\mathcal{N}(\tau)$ vanishes, meaning that the particle and the field are no longer affecting each other. In this case the equations decouple, apart from the holonomic constraint
\begin{equation}
    z(\tau) = u(x_\mathrm{p}(\tau)),\label{eq:holoO}
\end{equation} which defines a synchronization condition between the $u-$wave and the particle. This is reminiscent of the `phase harmony' condition introduced by de Broglie in his PhD thesis~\cite{debroglie1923}. This is fundamental in our model where the particle is guided by the $u-$wave.    
The dynamics of $z(\tau)$ then describes a simple harmonic motion
\begin{equation}
    z(\tau) = z_0 e^{-i\Omega_\mathrm{p}\tau},\label{eq:holo}
\end{equation}
while the particle with coordinate $x_\mathrm{p}(\tau)$ and the $u-$field follow the equations
\begin{align}
    m_\mathrm{eff.}\frac{d^2x_{\mathrm{p}\mu}(\tau)}{d\tau^2} &= e F_{\mu\nu}(x_\mathrm{p}(\tau))\dot{x}_\mathrm{p}^\nu\label{eq:motion_x}\\
    D^2u(x) &= 0\label{eq:motion_u}
\end{align}
with $m_\mathrm{eff.} = m_\mathrm{p}\pqty{1 + \sigma \Omega_\mathrm{p}^2\abs*{z_0}^2}$ an effective `dressed' mass that takes into account both the usual inertia $m_\mathrm{p}$ and the oscillatory motion along $z$, $F_{\mu\nu} = \partial_\mu A_\nu - \partial_\nu A_\mu$ the electromagnetic tensor and $D_\mu = \partial_\mu + i e A_\mu$ the covariant derivative. Eq.~\ref{eq:motion_x} describes the usual Lorentz force acting on a point-like particle (with dressed mass), and Eq.~\ref{eq:motion_u} describes a  field dynamics similar to the Klein-Gordon equation but without a mass term. The coupling between the wave and the particle is fixed by the holonomic constraint Eq.~\ref{eq:holoO} and the internal oscillation  Eq.~\ref{eq:holo}.

Solving Eqs.~\ref{eq:holo},\ref{eq:motion_x},\ref{eq:motion_u} may prove difficult because of both the relativistic formalism and the presence of the four-vector potential involving electric and magnetic contributions. As we showed in JD in presence of a Coulomb electrostatic field we can obtain a solution of Eq.~\ref{eq:motion_u} defining a guiding wave for the particle moving along a circular orbit.  The phase harmony condition $z=u$ forces the particle motion to be quantized and  the dynamics reproduces the Bohr-Sommerfeld quantization formula for the energy levels, velocities and radii of the particle. Inspired by de Broglie's work\cite{debroglie1925,debroglie1927,deBroglie1960} we found a solution of Eq.~\ref{eq:motion_u} where $u(x)$ is a sum of two counter-propagating modes with frequencies $\omega_\pm$ and orbital angular momentum $L_z:=m_\pm$ (i.e., projected along the $z-$axis). This solution induces a supraluminal phase-wave having all the properties of the wave introduced by de Broglie in his PhD work\cite{debroglie1925}. In turn, the particle-wave system fulfills the holonomic constraint $z=u$ along the orbit as it is required in order to induce the quantization  of the particle motion.   Here, we introduce an additional constant magnetic field and we will see that it is still possible to give approximate solutions in the non-relativistic and weak magnetic field limits corresponding to the standard Zeeman effect.

\section{Solutions of the equations of motion}
\label{sec:2}
\subsection{The particle dynamics}
\label{sec:2a}
Assuming a uniform circular motion in the $(x,y):=(\rho,\varphi)$ plane, let us write equation (\ref{eq:motion_x}) with the specific form of the four electromagnetic potential $A = (-\frac{e}{4\pi r}, \half B \rho \vu*{e}_\varphi)$:
\begin{equation}
    -m_\mathrm{eff.}\gamma\frac{v^2}{r} = -\frac{\alpha}{r^2} + e v B
    \label{eq:motion_p}
\end{equation}
 with $v$ the particle velocity and $\gamma=(1-v^2)^{-\frac{1}{2}}$ and where we used the Sommerfeld fine structure constant $\alpha=\frac{e^2}{4\pi}$. Eq.~\ref{eq:motion_p} describes the classical relativistic mechanics of a point-like particle along a circular orbit in a mixed electro and magneto static external field.  Similarly we can define the integral of motion associated with the angular orbital momentum by using the action variable formalism: 
\begin{equation}
    J:=\oint P\dd{l} = 2\pi r\pqty{\gamma m_\mathrm{eff.}v + \half e B r} = 2\pi n.
    \label{eq:quant_p}
\end{equation} This notation is similar to the one used by Sommerfeld in his old quantum mechanics\cite{Sommerfeld}. 
Note first, that for simplicity we identified the spherical radius $r$ with the cylindrical coordinate $\rho$, since we impose a magnetic field along the direction $z$ normal to the plane of motion $z = 0$. We also have introduced a number $n$ defining an integral of motion for the dynamics, i.e., the orbital angular momentum $L_z=n$ of the particle.  Knowing that, from now on we will index all physical quantities related to the particle's motion with this number $n$. Finally, we will introduce the Larmor frequency defined by $\omega_L = -eB/(2m_\mathrm{eff.})$, which will be derived later in this article using Larmor's theorem (see Sec. \ref{sec:2b}).

Eqs.~\ref{eq:motion_p} and \ref{eq:quant_p} allow us to derive the characteristic properties of the motion; in particular we can get the constant particle velocity along the orbit:
\begin{equation}
    v_n = \frac{\alpha}{n}\frac{1}{1 - \frac{m_\mathrm{eff.}\omega_L r_n^2}{n}}.
\end{equation}
In absence of magnetic field we recover the formula $v_n^{(0)} = \frac{\alpha}{n}$ for the purely Coulombian atom as obtained by Bohr\cite{Bohr1913} and justified in JD for our model.
We also have, in the non-relativistic limit, a fourth order equation for the radius $r_n$
\begin{equation}
    (m_\mathrm{eff.}\omega_L)^2r_n^4 + m_\mathrm{eff.}\alpha r_n - n^2 = 0.
\end{equation}
In the case of a weak magnetic field, we can find an approximate solution for this equation with a perturbation around the radius without magnetic field $r_n^{(0)} = n^2/(m_\mathrm{eff.}\alpha)$ (if $n=1$ we recover the standard Bohr radius $1/(m_\mathrm{eff.}\alpha)$). We get
\begin{equation}
    r_n \approx \frac{n^2}{m_\mathrm{eff.}\alpha}\pqty{1 - (m_\mathrm{eff.}\omega_L)^2\frac{n^6}{(m_\mathrm{eff.}\alpha)^4}}
\end{equation}
which will be equal to $r_n^{(0)}$ at first order in $B$. We can then use this expression in the velocity and we get
\begin{eqnarray}
    v_n \approx \frac{\alpha}{n}\pqty{1 + (m_\mathrm{eff.}\omega_L)\frac{n^3}{(m_\mathrm{eff.}\alpha)^2}}
    =\frac{\alpha}{n}+\omega_Lr_n^{(0)}.\label{eq:zeeman_vp}
\end{eqnarray}
This time the deviation from $v_n^{(0)} = \alpha/n$ is of order 1 in $B$. This formula is very intuitive and can recovered using Larmor's theorem by adding to  $v_n^{(0)}$ a velocity $\omega_Lr_n^{(0)}$ associated with a change of reference frame: This is explained in Section \ref{sec:2b}. \\
\indent In the end, the energy of the particle is
\begin{equation}
    E_n \approx m_\mathrm{eff.} + \half m_\mathrm{eff.}v_n^2 - \frac{\alpha}{r_n} \approx m_\mathrm{eff.}\pqty{1 - \half\frac{\alpha^2}{n^2}} + n \omega_L,
    \label{eq:zeeman_energy}
\end{equation}
and we recover the usual expression for the Zeeman splitting in the weak field and non-relativistic limits: \begin{equation}
E_n\approx E_n^{(0)}+n\omega_L=E_n^{(0)}-L_z\frac{eB}{2m_\mathrm{eff.}}.\label{eq:zeeman_energy_Bis}
\end{equation} where $E_n^{(0)}=m_\mathrm{eff.}\pqty{1 - \half\frac{\alpha^2}{n^2}}$ is the standard Bohr formula for the hydrogen-like atom energy (including a constant mass energy $m_\mathrm{eff.}$). 
Moreover, observe that the previous formulas are valid irrespectively of the sign of $B$ and $n$. If $n>0$ we have a positive orbital angular momentum $L_z=n$ and the orbit is followed in the anticlockwise direction,  whereas if  $n=-|n|<0$ we have a negative orbital angular momentum $L_z=-|n|<0$ and the orbit is followed in the clockwise direction. Therefore, for a given magnetic field along $z$, say $B>0$, we have two solutions for the  Zeeman effect corresponding to $E_n\approx E_n^{(0)}\pm|n|\omega_L$: The symmetry between the two travel directions along the orbit existing in absence of  $B$  has been broken. Of course, the same broken symmetry is obtained if instead of considering the two possible travel direction along the orbit we reverse the magnetic field so as to have $B=-|B|<0$. This equivalence will be used in Sec.~\ref{sec:2b} dealing with Larmor's theorem.\\ 
\indent At this point we stress that $n$ does not need to be an integer, i.e., we  still have a completely classical unquantized motion for the particle. Indeed, the Zeeman effect~\cite{Zeeman} is not inherently quantum mechanical but can also be found in classical mechanics as it was historically done by Lorentz~\cite{Lorentz} and Larmor~\cite{Larmor}. Moreover, in the following we will show that this Bohr-Sommerfeld quantization of $n$ is required as soon as we reintroduce the holonomic constraint with the field as was  shown previously in JD for the case of the pure Coulombian field.\\
\indent We also stress that Eq.~\ref{eq:zeeman_energy_Bis} is a particular case of the  formula $E_n=E_n^{(0)}-L_z\frac{eB}{2m_\mathrm{eff.}}$ where $n$ and $L_z$ (historically named the magnetic quantum number) are generally different.   Indeed, here we considered a motion in the $z=0$ plane where $L_z$ is identical to the full orbital angular momentum. In the more general case we can write $L_z=\cos{\theta_0}|\mathbf{L}|$  where $\theta_0$ is the angle between the orbital angular momentum pseudo vector $\mathbf{L}=m_\mathrm{eff.}\mathbf{x}\times\mathbf{v}$ and the $z$ axis oriented along the magnetic field. The effect of the magnetic field perturbation is to induce a slow precession of the electron orbital plane along the $z$-axis with Larmor's frequency $\omega_L$.  Here, we have only considered the case $\theta_0=0$ or $\pi$ corresponding to the anticlockwise and clockwise motions in the $z=0$ plane around the magnetic field axis. Moreover, we also supposed $|n|=|\mathbf{L}|$ associated with the simplest Bohr-like circular motion.  The most general case considered by Sommerfeld~\cite{Sommerfeld} allows $|n|\geq |\mathbf{L}|$ corresponding to elliptical orbits with excentricity $\epsilon=\sqrt{1-\frac{|\mathbf{L}|^2}{n^2}}$ (the case $n=0$ is excluded). Here, the restriction $\epsilon=0$ to a pure circular orbit is imposed by our theoretical model coupling the particle to a wave motion. More precisely, as we will show in Sec. \ref{sec:2b} the very demanding holonomic condition Eq.~\ref{eq:holo} requires the velocity to be a constant of motion which is not compatible with a general elliptical motion.   
\subsection{Larmor's theorem and the solutions for the $u$ field}
\label{sec:2b}
Solving the full wave equation for the field in presence of the electromagnetic potential
\begin{equation}
    \pqty{\partial_t - i  \frac{\alpha}{r}}^2u - \pqty{\grad - \half i e B \rho\vu*{e}_\varphi}^2u = 0 \label{rep}
\end{equation}
is not an easy task. It is however possible to find an approximate solution by emulating the magnetic field with an inertial Coriolis force, as is explained by Larmor's theorem.

\begin{figure}[h]
    \centering
    \includegraphics[width=0.6\linewidth]{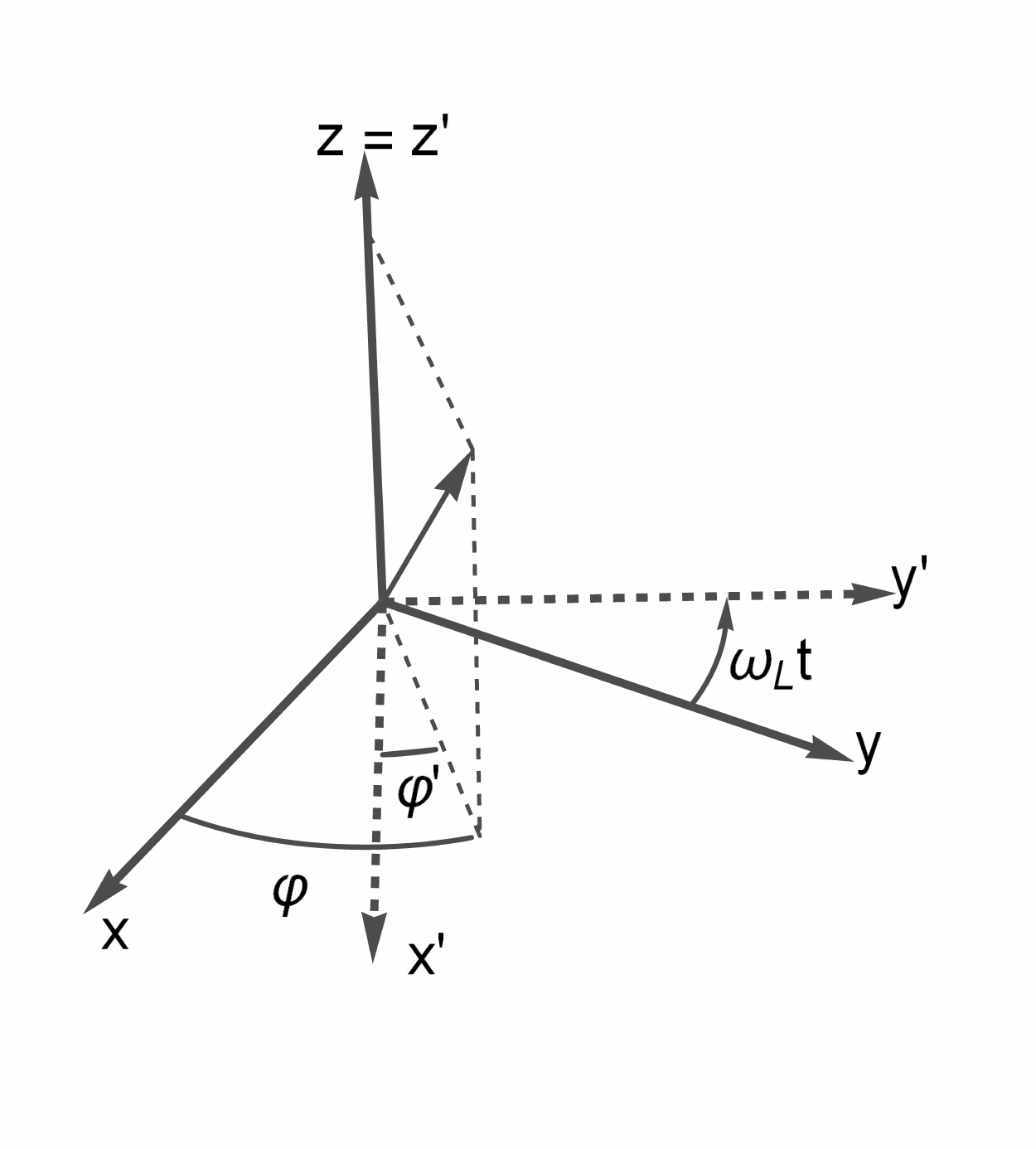}
    \caption{Illustration of the change of coordinates following Larmor's theorem. The new coordinates $x',y',z'$ (or equivalently $\rho',\varphi',z'$) are associated with a rotation of the original coordinate system $x,y,z$  (or equivalently $\rho,\varphi,z$) by an angle $\omega_Lt$ around the $z$ axis.}
    \label{fig:coordinate}
\end{figure}

\begin{figure}
    \centering
    \includegraphics[width=0.6\linewidth]{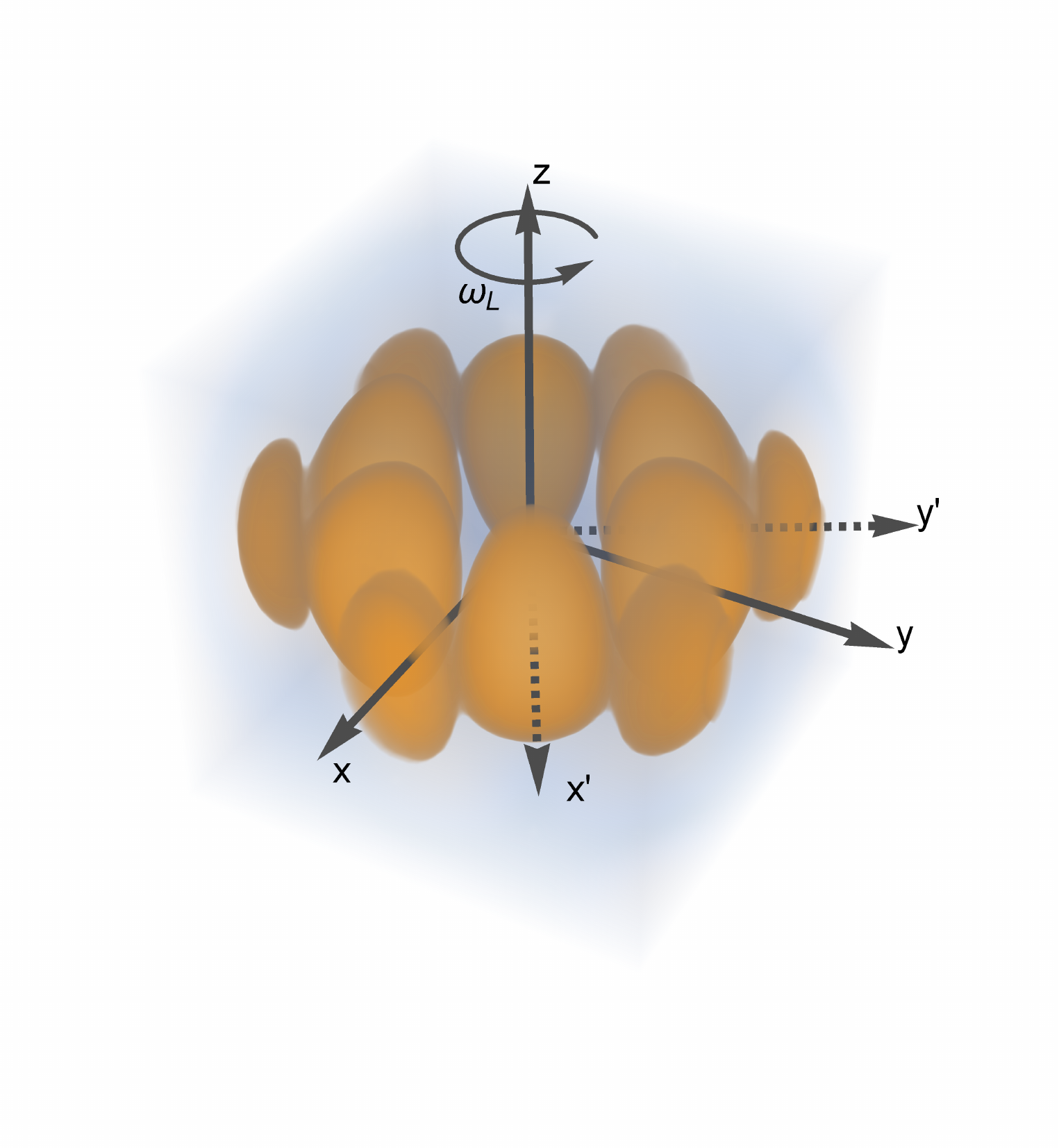}
    \caption{Rotation of the total $u-$field following Larmor's theorem. In color is represented the density of the modulus of the total field $\abs{u(0,\vb{x})}$ in the case $n = 1$, $m_+ = 4$, $m_- = 2$.}
    \label{fig:larmor}
\end{figure}

Consider first the following wave equation 
\begin{equation}
    \pqty{\partial_{t'} - i  \frac{\alpha}{r'}}^2u' - \pqty{\grad' }^2u' = 0 \label{temp}
\end{equation}
defined in an inertial reference frame $\mathcal{R}'$ where the scalar $u$-field reads $u'(t', \mathbf{x}')$. When this wave equation is watched from the point of view of a accelerated reference frame $\mathcal{R}$ rotating uniformly around the $z=z'$ axis this wave equation will be modified to include some inertial force terms.  For this purpose we use the following cylindrical coordinates transformation (sketched in Fig.~\ref{fig:coordinate})
\begin{align}
    t^\prime &= t \nonumber \\
    \rho^\prime &= \rho \nonumber\\
    z^\prime &= z\nonumber \\
    \varphi^\prime &= \varphi - \omega_L t \label{transf}
\end{align} 
where $t, \rho,\varphi, z$ are coordinates in the accelerated reference frame  and $\omega_L$ is the constant rotation frequency of $\mathcal{R'}$ relatively to $\mathcal{R}$ (the rotation is supposed non relativistic). 
This transformation in turn modifies the temporal derivative
\begin{equation}
	\partial_{t^\prime} =\partial_{t}  + \omega_L\partial_{\varphi^\prime}.
\end{equation} but otherwise lets the spatial derivatives unchanged: $\boldsymbol{\nabla}=\boldsymbol{\nabla}'$.\\ 
\indent Applying this transformation to Eq.~\ref{temp}  leads  (after some calculations detailed in Appendix \ref{app1}) to
\begin{equation}
	\pqty{\partial_{t} - i \frac{\alpha}{r}}^2u - \grad^2u - 2i\omega_L m_\mathrm{eff.}\partial_\varphi u + \mathcal{O}(\omega_L^2)\approx 0.\label{a}
\end{equation}
On the other hand, going back to Eq.~\ref{rep} the spatial operator can also be expanded
\begin{equation}
	 \pqty{\grad - \half i e B \rho \vu*{e}_\varphi}^2u = \grad^2u - i e B \partial_\varphi u + \mathcal{O}(B^2),
\end{equation}
and Eq.~\ref{rep} reduces to
\begin{equation}
 \pqty{\partial_{t} - i \frac{\alpha}{r}}^2u - \grad^2u  + i e B \partial_\varphi u + \mathcal{O}(B^2)\approx 0.\label{b}
\end{equation}
Comparing Eqs.~\ref{a} and \ref{b} we see that the two are identical (i.e., neglecting second order terms $\mathcal{O}(B^2)\sim \mathcal{O}(\omega_L^2)$) if  we carefully impose $\omega_L:=-e B/(2m_\mathrm{eff.})$ defining the Larmor frequency. In other words, a change of coordinates from the laboratory frame $\mathcal{R}$ to a new virtual frame $\mathcal{R'}$ rotating with the Larmor frequency $\omega_L$ exactly cancels out the effects of the magnetic field at the first order in $B$. This means that if we know the solution $u'$ without a magnetic field in a frame $\mathcal{R}^\prime$, we can get an approximate solution $u=u'$ in the laboratory frame $\mathcal{R}$ where a weak magnetic field is present by performing the rotation $\varphi = \varphi^\prime + \omega_L t^\prime$ as shown in Fig.~\ref{fig:coordinate}.\\
\indent In JD we previously obtained the solution of the wave equation $D^2u(x)=0$ in the presence of a Coulombian potential $eV(r) = -\alpha/r$.  In particular, we showed that in order to fulfill the holonomic condition $z(\tau)=u(x_n(\tau))$ the $u-$field must be written as $u(x)=u_+(x)+u_-(x)$ involving two counterpropagating modes $u_\pm$. This is necessary in order to obtain the correct de Broglie  `phase harmony' condition between the field and the oscillation $z$ of the particle. Moreover, along the circular orbit of radius $r_n$ we have
\begin{equation}
	u_\pm\pqty{t,r_n,\frac{\pi}{2},\varphi}  =\half u_0\mathrm{e}^{\mathrm{i}\pqty{\pm k_\pm r_n \varphi - \omega'_\pm t}},
\end{equation} where we introduced the two quantized wave vectors $k_\pm = \frac{m_\pm}{\pcl{r}}$ ($m_\pm \in \mathbb{N}$), and the mode frequencies $\omega'_\pm:=\omega_\pm^{(0)}$ (the complete expression for the field is given in JD and in Appendix~\ref{app2}). To go further we first write $u'_\pm(x')$  in the $\mathcal{R}^\prime$ frame where Eq.~\ref{temp} holds true. If we use Larmor's theorem in the laboratory frame $\mathcal{R}$ we get $u_\pm(t,r,\theta,\varphi)=u'_\pm(t',r',\theta',\varphi^\prime)$ defining our new solutions. Along the orbit we thus get:
\begin{equation}
	u_\pm\pqty{t,r_n,\frac{\pi}{2},\varphi}=\half u_0\mathrm{e}^{\mathrm{i}\pqty{\pm k_\pm r_n \varphi - (\omega_\pm^\prime \pm m_\pm\omega_L) t}}.
\end{equation} where appears the shifted frequencies $\omega_\pm^\prime \pm m_\pm\omega_L$. Note that in the first order approximation $r_n\simeq r_n^{(0)}$ as justified in Sec.~\ref{sec:2a} 
Moreover, along the orbit the total field $u(x)=u_+(x)+u_-(x)$ is
\begin{eqnarray}
u\pqty{t,r_n,\frac{\pi}{2},\varphi} = u_0 e^{i(\tilde{n}\varphi - (\omega^\prime_n + \tilde{n}\omega_L))t}\nonumber\\
\cdot\cos{\bqty{(\omega^\prime_n + \tilde{n}\omega_L+\varepsilon)\pqty{r_n \varphi - \frac{k_n - \eta}{\omega^\prime_n + \tilde{n}\omega_L + \varepsilon}t}}}\label{truc}
\end{eqnarray}
 with
\begin{equation}
    k_n = \frac{k_+ - k_-}{2},\quad\omega_n = \frac{\omega_+ + \omega_-}{2},\quad\omega'_n = \frac{\omega'_+ + \omega'_-}{2}
\end{equation} and where we introduced 
\begin{equation}
    \eta = \frac{\varepsilon_+ - \varepsilon_-}{2}~,\quad\varepsilon=\frac{\varepsilon_+ + \varepsilon_-}{2}.
\end{equation}
Moreover, the total field in the 3D space is obtained by summing the two modes and we illustrate in Fig.~\ref{fig:larmor} the case of a simple combination  of modes  $m_+ = 4$, $m_- = 2$ leading to the energy level $n = 1$.  The wave is rotating (precessing) with the frequency $\omega_L$ in agreement with Larmor's theorem.
Importantly, in Eq.~\ref{truc} we have inserted  a quantum number $\tilde{n} =\frac{m_+ - m_-}{2}\in \mathbb{Z}/2$. Thus, the mean frequency in the laboratory frame $\omega_n$ is related to the mean  frequency in the rotating frame $\omega_n^\prime$ by 
\begin{equation}
    \omega_n = \omega^\prime_n + \tilde{n}\omega_L= \omega^{(0)}_n + \tilde{n}\omega_L. \label{eq:zeeman_energy_tri}
\end{equation}
This looks like the expression Eq.~\ref{eq:zeeman_energy_Bis} for Zeeman's energy-shift when applying a weak magnetic field.\\
\indent In order for this analogy to become an identity   we must first have $\tilde{n}=n$. This is justified in our approach~\cite{Drezet2020,Jamet2021} if, in agreement with de Broglie~\cite{debroglie1923,debroglie1925}, we identify the group velocity of the $u-$wave in Eq.~\ref{truc}:
\begin{equation}
    v_\mathrm{g} = \frac{k - \eta}{\omega + \varepsilon}
\end{equation}
to the velocity of the particle along the circular trajectory:
\begin{equation}
    v_n = \frac{P_n - e A_\varphi}{E_n - e V},\label{vitesse}
\end{equation} with the canonical momentum $P_n=\gamma m_\mathrm{eff.}v_n + e A_\varphi=\gamma m_\mathrm{eff.}v_n + \half e B r_n$. The equality $v_n=v_g$ implies that the particle is constantly in phase with the guiding wave. With this condition the term $r_n \varphi - v_gt$ becomes a constant in Eq.~\ref{truc}. In other words, this corresponds to a synchronization of the internal clock of the particle with the phase of the $u-$wave as required by the holonomic condition $z=u$.  This motivates  the fundamental de Broglie relations: 
\begin{equation}
    P_n = b k_n~\mathrm{and}\quad E_n = b\omega_n\label{deb}
\end{equation}  with $b$ an apriori arbitrary constant. In our previous work JD we showed that a self consistent model is easily obtained if we impose  $b=1$ as a condition.   In the following we will thus accept the requirement $b=1$ for simplicity but we emphasize that in principle our model could be generalized to $b\neq 1$ as explained in JD.\\
\indent The condition $P_n=k_n=\frac{\tilde{n}}{r_n}$ is clearly reminiscent of Eq.\ref{eq:quant_p} which can be written
\begin{eqnarray}
\oint P_ndl= 2\pi r_n \oint k_n d\varphi=2\pi\tilde{n}.
\end{eqnarray}  This is indeed the Bohr-Sommerfeld quantization condition if $n=\tilde{n}$ as is also required to identify Eq.~\ref{eq:zeeman_energy_Bis} and Eq.~\ref{eq:zeeman_energy_tri}.    \\ 
\indent Moreover, the condition $v_n=v_g$ together with  Eq.~\ref{deb} lead to 
\begin{equation}
	\varepsilon_\pm = \frac{\alpha}{r_n} \pm \half e B r_n,
\end{equation}
and therefore from $k_\pm = \frac{n_\pm}{r_n} = \omega_\pm + \varepsilon_\pm$ we deduce
\begin{equation}
	E_n=\omega_n = \frac{N}{r_n} - \varepsilon=\frac{N}{r_n} - \frac{\alpha}{r_n}
\end{equation} where we defined $N=\frac{m_+ + m_-}{2}$.
Inserting $E_n$ 	and $P_n=n/r_n$ into Eq.~\ref{vitesse} we then easily find that
\begin{eqnarray}
    v_\mathrm{p} = v_\mathrm{g} \approx \frac{n}{N}\pqty{1 + \frac{m_\mathrm{eff.}\omega_L}{(m_\mathrm{eff.}\alpha)^2}n^3}+ \mathcal{O}(B^2)\nonumber\\
   =\frac{n}{N}\pqty{1 + \frac{m_\mathrm{eff.}\omega_Lr^{(0)}_n}{n}}+ \mathcal{O}(B^2)
\end{eqnarray}
from which we can deduce an interesting condition on the value of the fine structure constant $\alpha$ by comparing to Eq.~\ref{eq:zeeman_vp}:
\begin{equation}
    \alpha \approx \frac{n^2}{N} +\mathcal{O}(B^2) \equiv \alpha_0 + \mathcal{O}(B^2)
\end{equation}
where $\alpha_0 = n^2/N$ is the value for an isolated atom, \textit{i.e.} without a magnetic field as studied in JD. This relation is fundamental since it provides a selection rule for the different possible values of  $n$ and $N$ or equivalently $m_\pm$.  As we showed in JD if $\alpha_0^{-1}$ is an integer (e.g., $\alpha_0^{-1}=137$) then the allowed values for $n$ are necessarily limited to integer numbers $|n|=1,2...$. As justified in JD this is central to recover the Bohr-Sommerfeld quantization postulate since   a priori our number $n\equiv\tilde{n} =\frac{m_+ - m_-}{2}\in \mathbb{Z}/2$ is not necessarily an integer: Without this selection rule it would also allow half-integers $1/2,3/2...$. While half-quanta are not a priori useless in connection with fermionic spin (that was the initial reasonning of Heisenberg as a student of Sommerfeld~\cite{Sommerfeld}) we  prefer here to be rather conservative and follow the usual approach without half-quanta (reserving the possibility to come back to this issue in a future work).       

One final important relation  is obtained  by writing the holonomic constraint
\begin{equation}
    z(t) = z_0e^{-i\Omega_\mathrm{p}\sqrt{1 - v_n^2}t} = u(t,x_\mathrm{p}(t)) = u_0 e^{i\mathcal{L}_\mathrm{p}t} \label{eq:lagrangian_pBis}
\end{equation}
where $\mathcal{L}_\mathrm{p}$ is the particle Lagrangian
\begin{eqnarray}
    \mathcal{L}_\mathrm{p} = -m_\mathrm{eff.}\sqrt{1-v_n^2} - eV(r_n) +eA_\varphi(r_n)v_n \nonumber \\= -m_\mathrm{eff.}\sqrt{1-v_n^2} + \frac{\alpha}{r_n} + \half e B r_n v_n
    \label{eq:lagrangian_p}
\end{eqnarray}
which is of course the  classical Lagrangian for a relativistic particle in an electromagnetic field. The presence of the Lagrangian comes from the fact that the phase  of $u(t,x_\mathrm{p}(t))$ can be written $(k_nv_n-\omega_n)t=(P_nv_n-E_n)t$ which by definition is $\mathcal{L}_\mathrm{p}t$. Therefore,  as required by the holonomic condition we have equal amplitudes $z_0 = u_0$ as well as equal phases:
\begin{eqnarray}
    \Omega_\mathrm{p} = m_\mathrm{eff.} - \pqty{\frac{\alpha}{r_n} +\half e B r_n v_n}\frac{1}{\sqrt{1 - v_n^2}}\nonumber\\ \approx m_{\mathrm{eff.}}(1-\frac{\alpha^2}{n^2}) +\omega_Ln(1-\frac{\alpha^2}{2n^2}),
\end{eqnarray} 
which, remembering that $m_\mathrm{eff.} = m_\mathrm{p}(1 + \sigma \Omega_\mathrm{p}^2\abs{z_0}^2)$, completely determine the amplitude $z_0$ and thus all the characteristics of the motion using the different parameters of our model.\\
\indent Before concluding, a few points must be emphasized in order to show the consistency of our model: i) First, in JD the frequencies $\omega_\pm^{(0)} $ of the modes were calculated to be:  \begin{eqnarray}
     \omega_\pm^{(0)}=\frac{m_{\mathrm{eff.}}}{\sqrt{1 - \frac{\alpha^2}{n^2}}} (1\pm\frac{\alpha}{n}-\frac{\alpha^2}{n^2})\nonumber\\
     \approx m_{\mathrm{eff.}}(1\pm\frac{\alpha}{n}-\frac{\alpha^2}{2n^2}).
	\label{eq:w_vec_freqtri}
\end{eqnarray} The dominant contribution is the effective mass $m_{\mathrm{eff.}}$ and this is not modified if we include the first order magnetic perturbation so that $\omega_\pm=\omega_\pm^{(0)} \pm m_\pm\omega_L\sim m_{\mathrm{eff.}}$. This plays a role in self consistency of our method (see Appendix \ref{app1}).\\
\indent ii) Second, we stress that the Larmor theorem is usually presented for particles and not for waves. Moreover, the derivation of the particle dynamics given in Sec.~\ref{sec:2a} could be done directly using the Larmor theorem as it was proposed by Larmor himself~\cite{Larmor}. The derivation is straightforward: Starting from the inertial reference frame $\mathcal{R}'$ where the nonrelativistic particle dynamics is described by a Newtonian equation 
\begin{equation}
m_\mathrm{eff.}\frac{d}{dt'}\mathbf{v}'_p(t')=-\frac{\alpha\mathbf{\hat{r}}'}{r'^2}
\end{equation} that involves a Coulombian (electrostatic) force, and using the coordinate transformation  Eq.~\ref{transf} to the accelerated reference frame $\mathcal{R}$ uniformly rotating around the $z$ axis with an angular velocity $\omega_L\mathbf{\hat{z}}$ we obtain:
\begin{eqnarray}
m_\mathrm{eff.}\frac{d}{dt}\mathbf{v}_p(t)\approx-\frac{\alpha\mathbf{\hat{r}}}{r^2} -2m_\mathrm{eff.}\mathbf{v}_p(t)\times\omega_L\mathbf{\hat{z}}+ \mathcal{O}(\omega_L^2).
\end{eqnarray}  Up to second order terms in $\omega_L$ this equation is similar to the dynamics of a particle submitted to both a central Coulombian force and to the Laplace-Ampere-Lorentz force in a constant magnetic field $\mathbf{B}=B\mathbf{\hat{z}}$, i.e.,
   \begin{eqnarray}
m_\mathrm{eff.}\frac{d}{dt}\mathbf{v}_p(t)=-\frac{\alpha\mathbf{\hat{r}}}{r^2} +e\mathbf{v}_p(t)\times B\mathbf{\hat{z}}.
\end{eqnarray} 
The identity occurs if we assume $\omega_L=-\frac{eB}{2m_\mathrm{eff.}}$ that constitutes the Larmor formula. Importantly, the particle velocity in the $\mathcal{R}$ reference frame is given by $\mathbf{v}_p(t)=\mathbf{v}'_p(t')+\omega_L\mathbf{\hat{z}}\times\mathbf{r}$, i.e., 
\begin{eqnarray}
v_\varphi=v'_\varphi+\omega_Lr:=\frac{\alpha}{n}+\omega_Lr_n
\end{eqnarray} which is equivalent to Eq.~\ref{eq:zeeman_vp} $v=\frac{\alpha}{n}+\omega_Lr_n^{(0)}$ with $r_n\simeq r_n^{(0)}$. In the end, the nonrelativistic particle energy 
$E'-m_\mathrm{eff.}=\frac{1}{2}m_\mathrm{eff.}{v'}^2-\frac{\alpha}{r'}$ in $\mathcal{R'}$ transforms into the energy  $E-m_\mathrm{eff.}=\frac{1}{2}m_\mathrm{eff.}v^2-\frac{\alpha}{r}= E'+m_\mathrm{eff.}\omega_Lr_nv_n +\mathcal{O}(\omega_L^2)$ in $\mathcal{R'}$.  Using the formula  for $r_n\simeq n^2/{m_\mathrm{eff.}\alpha}$ and $v_n$ leads to 
\begin{eqnarray}
E_n\simeq E'_n+n\omega_L +\mathcal{O}(\omega_L^2)
\end{eqnarray} recovering the Zeeman splitting of Eq.~\ref{eq:zeeman_energy_Bis}. Therefore, we see that the two applications of the Larmor theorem to respectively the wave and the particle are self consistent. The coordinate transformation  Eq.~\ref{transf} from the reference frame $\mathcal{R}'$ to $\mathcal{R}$ leads to a circular particle orbit with a modified velocity $v_n=\frac{\alpha}{n}+\omega_Lr_n^{(0)}$ whereas the $u-$field is also modified accordingly to obey Eq.~\ref{truc}. In the $\mathcal{R}'$ reference frame the $u-$field is given by the unperturbed  wave field associated with the pure Coulombian solution studied in JD.  The Larmor theorem applied to both the field and the particle allows us to obtain the new dynamics in presence of a weak constant magnetic field $B$ valid in the laboratory reference frame $\mathcal{R}$.\\   
\indent iii) As a final remark, we must return to  our comment at the end of Sec.~\ref{sec:2a} concerning the limitation of our model to uniform circular motion.   For this purpose it is enough to consider the regime without external magnetic field, i.e., $B=0$, and to use once more the holonomic condition Eq.~\ref{eq:lagrangian_pBis}  with the Lagrangian given by Eq.~\ref{eq:lagrangian_p}.   This leads to  
\begin{eqnarray}
     (\Omega_\mathrm{p}-m_\mathrm{eff.})\gamma_n^{-1}=-\alpha/r_n=E_n-m_\mathrm{eff.}\gamma_n\label{veloc}
\end{eqnarray}  where  we used the definition of the constant energy $E_n=m_\mathrm{eff.}\gamma_n-\alpha/r_n$  (with $\gamma_n=1/\sqrt{1-v_n^2}$).   
Solving Eq.~\ref{veloc} gives  \begin{eqnarray}
\gamma_n=\frac{E_n}{2m_\mathrm{eff.}}+\frac{1}{2m_\mathrm{eff.}}\sqrt{(E_n^2-4m_\mathrm{eff.}(\Omega_\mathrm{p}-m_\mathrm{eff.}))},
\end{eqnarray} that corresponds to a constant velocity. This clearly shows the intrinsic limitation of our model associated with the constant value of the internal frequency $\Omega_\mathrm{p}$.

\section{Discussion and perspectives}
In this manuscript, we have extended our 3D mechanical atomic model presented in JD by adding an external uniform magnetic field. We recovered the mathematics of the normal Zeeman splitting of the energy levels in the atom in the weak field limit by making use of an equivalence between a constant and uniform magnetic field and inertial forces of a rotating frame of reference. 

Going back to the results of the present work, we showed that the quantum Zeeman effect arises naturally in our model. This extends the regime of applicability of the analogy discussed  in our previous article JD. There are however some limitations that need to be considered, mainly the difficulties of our model to introduce elliptical orbits, or the case of $s$-orbitals with $n = 0$ that are part of quantum mechanics.
Moreover,  here we limited our analysis to the  transparency regime where $\mathcal{N}=0$. Going beyond  would imply to consider transitions between orbits (e.g., to study the stability  of trajectories or the chaotic nature of the dynamics). Such a regime clearly needs further considerations. 

In JD we suggested ways of experimentally reproducing the Bohr-Sommerfeld quantum condition Eq.~\ref{eq:quant_p} as implementations of our atomic model, in particular using optical vortices with definite orbital angular momenta to trap small particles on quantized orbits (following the method developped by Ashkin \cite{Ashkin1970} for optical tweezers). This same system could naturally be extended to include a transverse magnetic field which, by interacting with the orbital angular momentum, should also reproduce the results shown in this paper, i.e. the modification of all orbital quantities following Zeeman's results. The most important feature in the optical tweezers analogy is the holonomic condition  $z(\tau) = u(x_\mathrm{p}(\tau))$ that can be seen as particular weak resonance condition for a nano-particle or nano-antenna  in a optical~\cite{Crozier2019} or plasmonic field~\cite{Cuche2012} (that is if we can neglect scattering). Further studies should be done in order to understand this interesting  analogy in both the 2D and 3D regime. Potential applications in biology or micro/nano technology could ultimately be considered.  More generally, if one were to produce an experimental demonstration of this system, introducing inertial forces would suffice at first order to mimic the effects of an electromagnetic field, as was hinted in other experimental and theoretical works on hydrodynamical analogs\cite{Fort2010,Harris1991}. This opens interesting perspectives for discussing analogies with the Zeeman effect using mechanical or acoustical systems. This was the original motivation of our nonrelativistic model~\cite{Drezet2020} where a particle sliding on a vibrating string was guided by the phase wave generated by two counter propagating plane waves $u_\pm$.  Clearly, this shows the strong transdisciplinarity of  our analogy and method. Furthermore,  the different mathematical tools used in our model coupling a wave to a guided particle also stress the links betwen methods applied to classical mechanics of point-like objects and the dynamics of waves (in full agreement with the goal of de Broglie in his double solution theory\cite{deBroglie1960}).   We believe that this constitutes a strength of our general methodology that can be applied to different fields.\\     
\indent Moreover, at a fundamental level it should be noted that this equivalence between a constant and uniform magnetic field and inertial forces of a rotating frame of reference could be taken one step further by using a general relativity formalism and interpreting the whole electromagnetic field as a curvature in the metric. The idea is to find some kind of an equivalence between the two covariant derivatives in general relativity and our formalism, 
\begin{equation}
    \partial_\mu + i e A_\mu \sim \partial_\mu + \tilde{\Gamma}_{\mu}
    \label{eq:equiv_GR_elm}
\end{equation}
with $\tilde{\Gamma}_\mu$ some contraction or operation on the Levi-Civita connection $\Gamma_{\mu\,\nu}^{\;\rho}$ defined in general relativity by 
\begin{equation}
    \Gamma_{\mu\,\nu}^{\;\rho} = \half g^{\rho\lambda}\pqty{\partial_\nu g_{\mu\lambda} + \partial_\mu g_{\nu\lambda} - \partial_\lambda g_{\mu\nu}},
\end{equation}
with $g_{\mu\nu}$ the metric tensor and $\partial_\mu$ the derivative with respect to the coordinate $x^\mu$.
This of course can not be done in the general case, and the equivalence~\ref{eq:equiv_GR_elm} is only here to represent schematically the goal of the analogy. However it was done for example in the case of weak static fields\cite{Harris1991,Semon1980} -- which is our case of interest -- using the electromagnetic tensor $F_{\mu\nu}$ and part of the affine connection $\Gamma_{0\,\nu}^{\;\rho}$
\begin{equation}
    \begin{aligned}
        F_{\mu\nu} &= \partial_\mu A_\nu - \partial_\nu A_\mu\\ 
        &\equiv \eta_{\mu\rho}\Gamma_{0\,\nu}^{\;\rho} = \half\eta_{\mu\rho}g^{\rho\lambda}\pqty{\partial_0 g_{\lambda\nu} + \partial_\nu g_{0\lambda} - \partial_\lambda g_{0\nu}}
    \end{aligned}
\end{equation}
with $\eta_{\mu\nu} = \mathrm{diag}(1,-1,-1,-1)$ the Minkowski metric.
On top of being interesting analogies in and of themselves, these ideas of emulating or equating `real' forces with inertial ones can prove computationnaly useful as was shown here by our solving of the covariant wave equation. This general relativity formalism has of course the added benefit of being able to account for both electric and magnetic forces with a combinaison of Coriolis and centrifugal forces.

In the end, the formalism developed in the present work allows us to consider very general cases of wave-particle duality with various external potentials such as a central electrostatic potential or a uniform magnetic field, with the two main assumptions, i.e. the existence of a periodic phenomenon in the particle and its holonomic coupling with the field, being at the root of all quantum phenomena. These are ideas that clearly deserve to be studied extensively and using various approaches (e.g. pilot wave theories, double solution theories, bohmian mechanics, hydrodynamic analogs\dots  \cite{Valentini2009, Bohm1952,Bohm1993,deBroglie1960}), as they are able to provide us with insight into the nature of quantum mechanics.
\appendix
\section{Larmor's theorem for the wave equation}\label{app1}
\indent The application of the tranformation Eq.~\ref{transf} to  Eq.~\ref{temp}  leads to:
\begin{eqnarray}
	\pqty{\partial_{t} - i \frac{\alpha}{r'}}^2 = \pqty{\partial_{t} - i \frac{\alpha}{r} + \omega_L\partial_\varphi}^2 \nonumber\\
	= \pqty{\partial_{t} - i \frac{\alpha}{r}}^2 + 2\omega_L\partial_{\varphi}\pqty{\partial_{t} - i \frac{\alpha}{r}} + \mathcal{O}(\omega_L^2)
\end{eqnarray} where we used $r=\sqrt{\rho^2+z^2}= \sqrt{\rho'^2+z'^2}=r^\prime$. Moreover, assuming a harmonic solution  for the u-field $u'(t',\vb{x'}):=u(t,\vb{x}) = f(\vb{x})e^{-i\omega t}$, and if we neglect the radial term $\alpha/r \ll \omega$, we can write
\begin{equation}
	\pqty{\partial_{t} - i\frac{\alpha}{r}}f(\vb{x})e^{-i\omega t}\approx -i\omega f(\vb{x})e^{-i\omega t},\label{oki}
\end{equation}
which finally gives us for the temporal part of the d'Alembert operator when we identify the energy $\omega$ with an equivalent mass $m_\mathrm{eff.}$:
\begin{equation}
	\pqty{\partial_{t} - i\frac{\alpha}{r} + \omega_L\partial_\varphi}^2 \approx \pqty{\partial_{t} - i \frac{\alpha}{r}}^2u - 2i\omega_L m_\mathrm{eff.}\partial_\varphi u + \mathcal{O}(\omega_L^2).\label{nezo}
\end{equation}
Moreover, we justify the identification  $\omega\sim m_\mathrm{eff.}$ used in Eq.~\ref{oki} afterwards. Indeed, from Eq.~\ref{eq:w_vec_freqtri}
we see that  the dominant contribution in the mode frequencies $\omega_\pm^{(0)}$ is the effective mass $m_{\mathrm{eff.}}$ and this is not modified if we include the first order magnetic perturbation so that $\omega_\pm=\omega_\pm^{(0)} \pm m_\pm\omega_L\sim m_{\mathrm{eff.}}$. This in turn justifies why it was judicious in Eq.~\ref{nezo} to identify the frequency $\omega$ with the mass term $m_{\mathrm{eff.}}$ (a similar method is used for deriving  Schr\"odinger's equation from Klein-Gordon's one). Therefore, the self consistency of our derivation of Larmor's theorem for the wave equation and of the related dispersion relation given by Eq.~\ref{zeemb} are a posteriori justified.
\section{Solutions of the wave equation}\label{app2}
\indent For the present problem writing the two solutions $u'_\pm(x')$  in the $\mathcal{R}^\prime$ frame where Eq.~\ref{temp} holds true we have 
(using spherical coordinates):
\begin{equation}
    u'_\pm(t',r',\theta',\varphi^\prime) = A_\pm R_{\tilde{l}_\pm}(r') P_{l_\pm}^{\pm m_\pm}(\cos\theta')e^{i(\pm m_\pm \varphi^\prime - \omega_\pm^\prime t)},
\end{equation} where $l_\pm\geq 0$ and $m_\pm\leq l_\pm$ are positive integers labeling spherical harmonics $P_{l_\pm}^{\pm m_\pm}(\cos\theta')e^{\pm i m_\pm \varphi^\prime}$, and where the radial contribution $R_{l'_\pm}(r')$  (with $\tilde{l}_\pm=\frac{-1}{2}+\sqrt{((l_\pm+\frac{1}{2})^2-\alpha^2)}$) is a complex function:
\begin{eqnarray}
R_{\tilde{l}_\pm}(r')=e^{i\omega'_\pm r'}r'^{\tilde{l}_\pm}M(\tilde{l}_\pm+1-i\alpha,2\tilde{l}_\pm+2,-2i\omega_\pm r')
\end{eqnarray}  defined using the Kummer confluent hypergeometric function $M(a,b,c)={}_1F_1(a,b,c)$.\\
\indent Moreover, using Larmor's theorem in the laboratory frame $\mathcal{R}$ we get $    u_\pm(t,r,\theta,\varphi)=u'_\pm(t',r',\theta',\varphi^\prime)$:
\begin{equation}
    u_\pm(t,r,\theta,\varphi) = A_\pm R_{\tilde{l}_\pm}(r) P_{l_\pm}^{\pm m_\pm}(\cos\theta)e^{i(\pm m_\pm \varphi - (\omega_\pm^\prime \pm m_\pm\omega_L) t)}.
\end{equation} This equation clearly shows that the mode frequencies $\omega_\pm$ in the $\mathcal{R}$ reference frame  are, according to Larmor's theorem, given by the formula
\begin{eqnarray}
\omega_\pm=\omega_\pm^\prime \pm m_\pm\omega_L\label{zeem}
\end{eqnarray} that is actually reminiscent of the Zeeman spectrum given in Eq.~\ref{eq:zeeman_energy_Bis}.  Therefore, in Eq.~\ref{zeem}  $\omega_\pm^\prime=\omega_\pm^{(0)}$ can be interpreted as the  zero-order solution in absence of magnetic field:
\begin{eqnarray}
\omega_\pm=\omega_\pm^{(0)} \pm m_\pm\omega_L\label{zeemb}
\end{eqnarray} 
\indent To go further, we recall from our previous article JD that along the circular path of the particle the two modes $u_\pm$ read 
\begin{equation}
	u_\pm\pqty{t,r_n,\frac{\pi}{2},\varphi}  =\half u_0\mathrm{e}^{\mathrm{i}\pqty{\pm k_\pm r_n \varphi - \omega_\pm t}},
\end{equation}
with the wave vectors condition $k_\pm = \frac{m_\pm}{\pcl{r}}$ ($m_\pm \in \mathbb{N}$). By definition of the modes $u_\pm$ we have $A_\pm R_{\tilde{l}_\pm}(r_n) P_{l_\pm}^{\pm m_\pm}(0)=\frac{u_0}{2}$ imposing a specific condition on the amplitudes of the two modes $u_+$ and $u_-$.   
\section*{Data Availability Statement}

Data sharing is not applicable to this article as no new data were created or analyzed in this study.

\bibliography{bibliography2022}

\end{document}